\documentclass[aps,prl, twocolumn,superscriptaddress,showpacs]{revtex4}
\usepackage{amsmath}
\usepackage{graphicx}

\usepackage[german,english]{babel}
\begin{document}

\title{Absence of Wigner Crystallization in Graphene.}

\author{Hari P. Dahal}
\affiliation{Department of Physics, Boston College, Chestnut Hill,
 MA, 02467}
 \affiliation{Theoretical Division, Los Alamos National Laboratory,
Los Alamos, New Maxico 87545}

\author{Yogesh N. Joglekar}
\affiliation{Department of Physics, Indiana University-Purdue
University Indianapolis, Indianapolis, Indiana 46202}
\affiliation{Theoretical Division, Los Alamos National Laboratory,
Los Alamos, New Maxico 87545}

\author{Kevin S. Bedell}
\affiliation{Department of Physics, Boston College, Chestnut Hill,
 MA, 02467}

\author{Alexander V. Balatsky}
\affiliation{Theoretical Division and Center for Integrated
Nanotechnology, Los Alamos National Laboratory, Los Alamos, New
Maxico 87545}
\email[]{* avb@lanl.gov, http://theory.lanl.gov}

\date{\today}

\begin{abstract}
Graphene, a single sheet of graphite, has attracted tremendous attention due 
to recent experiments which demonstrate that carriers in it are described 
by massless fermions with linear dispersion. In this note, we consider the 
possibility of Wigner crystallization in graphene in the absence of 
external magnetic field. We show that the ratio of potential and kinetic 
     energy is independent of the carrier density, the tuning parameter that 
usually drives Wigner crystallization and find that for given material 
parameters (dielectric constant and Fermi velocity), Wigner crystallization 
is not possible. We comment on the how these results change in the presence 
of a strong external magnetic field. 

\end{abstract}

\maketitle

Graphene is a single sheet of carbon atoms which can be extracted from 
graphite by micro-mechanical cleavage \cite{novoselov2005}. Various 
theoretical and experimental studies of this 2-dimensional system have 
shown that its properties
 \cite{zheng2002,gusynin2005,peres2006,novoselovnature, katsnelsonnature, novoselovnature2} are 
markedly different from those of the conventional 2-dimensional electron 
gas (2DEG) formed in semiconductor heterostructures. The carriers in such 
a 2DEG are well described by Fermi liquid theory, and have quadratic 
dispersion at small momenta, $E_k=\hbar^2k^2/2m^{*}$, where $m^{*}$ is the 
band mass \cite{ando1982}. Carriers in graphene, on the 
other hand, have a band structure in which electron and hole bands touch at 
two points in the Broullion zone. This, combined with the hexagonal lattice 
structure of graphene, lead to linear dispersion at small momenta, 
$E_k=\hbar v_G k$. Therefore 
the carriers behave as ``massless'' Dirac particles  \cite{wallace1947,semenoff1984,haldane1988,novoselovnature, katsnelsonnature} with characteristic 
velocity $v_G\sim 10^6$ m/s. 
This difference in the character of quasiparticles - massive quadratically 
dispersing vs. massless linearly dispersing - gives rise to a host of 
remarkable phenomena including unusual quantum Hall effects and $\pi$ 
Berry phase \cite{zheng2002,gusynin2005,novoselovnature,peres2006}. 

Another interesting aspect of graphene is that it is a better realization of 
two dimensional system. The conventional 2DEG is formed in semiconductor 
quantum wells where the width of the quantum well is usually around 100 - 300 
\AA. Various properties of such a 2DEG have been extensively studied using 
primarily transport measurements. However, it has not been amenable to local 
probes such as a scanning tunneling microscope (STM), because the 2DEG is 
buried about $1000$ $\AA$  from the sample surface. In graphene, on the other 
hand, the width of the effective quantum well is approximately $5 - 10$ $\AA$ 
(distance between two graphene sheets in graphite), and the 2DEG is amenable 
to local probes including those which will allow the study of inhomogeneous 
states. In addition, the density and the polarity of carriers in graphene
can be adjusted simply by changing the gate voltage; something that is not
possible in conventional semiconductor heterojunctions.

Here, we investigate the possibility of Wigner crystallization in graphene. 
Following the pioneering work of E. Wigner \cite{wigner1934}, a Wigner 
crystal has become one of the extensively studied phases of conventional 
2DEG at low densities. This state appears when electrons localize and form a 
crystal to minimize the potential energy, while paying the concommitant 
kinetic energy cost which arises from localization, as the density of 
carriers is lowered. Theoretical studies predict that Wigner crystallization 
in conventional 2DEG occurs at $r_s\sim 37$, where $r_sa_B$ is the mean 
interparticle spacing and $a_B$ is the Bohr radius \cite{tanatar1989}. 
In the presence of a strong magnetic field, the kinetic energy of electrons 
is quenched at filling factors $\nu\leq 1$; in other words, electrons can 
localize without paying the kinetic energy cost. Therefore, Wigner 
crystallization is facilitated by the magnetic field. The experimental study 
of the quantum hall effect in conventional 2DEG has shown the existence of 
the Wigner crystal phase at filling factors $\nu\leq 1/5$, deduced from 
vanishing Hall and longitudinal conductivity \cite{willett1988}. 

The situation in case of graphene is not yet fully explored. In the absence 
of a magnetic field, graphene shows metallic behavior over a large range of 
density for both polarities. In the quantum Hall regime it shows finite Hall 
and longitudinal conductivity, although only integer filling factors have 
been explored. Hence, it is appropriate to ask if Wigner crystallization 
occurs in graphene. In this note we show that {\it Wigner crystal phase 
is not possible in graphene except in the presence of an external magnetic 
field.}

First, we present a heuristic argument for our result. Our discussion is 
limited to low temperatures $T\rightarrow 0$; at finite temperatures, a 
translationally broken symmetry state will be destroyed by fluctuations in 
two dimensions \cite{mermin}. We know that the
Wigner crystal phase is formed as a result of the competition between the 
potential energy $E_p$ and kinetic energy $E_k$ of the system. When 
$E_p\gg E_k$ the crystallization occurs. We estimate the two energies in 
graphene as follows. Assuming localization of carriers at length scale $l$ 
(which, in turn, is related to the density of carriers), the potential 
energy is given by $E_p\sim (e^2/\epsilon l)nA$ where $n$ is the density 
of carriers, $A$ is the area of the system, and $\epsilon$ is the static 
dielectic constant of graphene. The kinetic energy, on the other hand, will 
be given by $E_k\sim\hbar v_G(2\pi/l)nA$ where $\hbar v_G=5.8$ 
$eV\AA$ is the velocity of massless Dirac fermions. Hence, the ratio of these 
two energies is 
\begin{equation}
\frac{E_p}{E_k}\sim \frac{e^2}{h} \frac{1}{\epsilon v_G}.
\end{equation}
Two important observations thus follow immediately: i) The ratio is 
independent of the density of carriers, a situation very different from 
that in the conventional 2DEG ii) The ratio depends on only two (tunable) 
material parameters, $\epsilon$ and $v_G$. 

Thus, for graphene, we find that energetics of the system do not depend on 
the carrier density. In particular, since it is known experimentally that 
graphene is not an insulator, it will not undergo Wigner crystallization 
to an insulator by reducing density of carriers. In the following, we 
present a microscopic calculation of the two energies, and obtain numerical 
estimates for suppressing the dielectric constant to increase the ratio 
$E_p/E_k$, which may drive the system towards a Wigner crystal. It is 
worthwhile to remember at this point that in conventional 2DEG, Wigner 
crystallization occurs for $E_p/E_k=r_s\sim 37$.

Now we will proceed with more detailed estimates of the kinetic and 
potential energy. In the ground state 
$|F\rangle$, the kinetic energy of the system is given by 
\begin{equation}
\begin{split}
E_k =\langle F|\widehat{H}_0|F\rangle & = \hbar v_G
\sum_{{\bf k}\lambda}k \langle F|a^{\dagger}_{{\bf k}\lambda}a_{{\bf k}
\lambda}|F\rangle\\
=\frac{\hbar v_G}{2\pi}\frac{4Ak_F^3}{3} 
\end{split}
\end{equation}
where $\sum_\lambda = 4$ is the sum over the spin and valley degeneracy, 
$a^{\dagger}_{{\bf k}\lambda}$ creates an electron with momentum ${\bf k}$ 
(measured from the Dirac point), 
and the Fermi momentum $k_F$ is related to the carrier density by 
$n=k_F^2/\pi$. Therefore, the kinetic energy becomes 
\begin{equation}
E_k=\frac{2\sqrt{\pi}}{3}\hbar v_G A n^{3/2}.
\end{equation}

Similarly the potential energy is calculated as follows
\begin{equation}
\begin{split}
E_p = \frac{e^2}{2A\epsilon}\sum_{{\bf k}{\bf p}{\bf q}}
\sum_{\lambda_1\lambda_2}\frac{2\pi}{q}
\langle F|a^{\dagger}_{{\bf k}+{\bf q},\lambda_1}a^{\dagger}_
{{\bf p}-{\bf q},\lambda_2}a_{{\bf p},\lambda_2}a_{{\bf k},\lambda_1}|F\rangle
\\
=\frac{4\sqrt{\pi} e^2}{3\epsilon}A n^{3/2}
\end{split}
\end{equation}

So their ratio is given by 
\begin{equation}
\frac{E_p}{E_k}= \frac{4e^2}{h}\frac{1}{\epsilon v_G}
\end{equation}
and is independent of density of carriers as expected. For the experimental 
parameters, $\epsilon=4.7$ and $\hbar v_G=5.8$ eV\AA, the ratio 
$E_p/E_k\simeq 0.32$. Thus, kinetic energy is roughly three times bigger 
than the potential energy, and the carriers prefer to be delocalized in a 
uniform state. For Wigner crystallization, we need $E_p/E_k\gg 1$. 

Next we consider the dependence of this ratio on the two tunable parameters, 
$\epsilon$ and $v_G$. To this end, we first calculate the change in the 
static dielectric constant, $\epsilon(q\rightarrow 0,\omega=0)$ due to the 
spin-orbit coupling $\Delta_{so}$ in graphene \cite{wang2006}. The spin-orbit 
coupling in graphene is estimated to be as high as 4 meV \cite{yao2006}. 
Therefore, following the derivationn in Ref.\onlinecite{wang2006}, 
we calculate the dielectric constant for $0\leq \Delta_{so}\leq 5$ meV.

At $\omega=0$, the dielectric functional $\epsilon(q)$ can be written as 
\begin{equation}
\epsilon({\bf q})= 1 + U_0(q)\Pi({\bf q})\end{equation}
where $U_0(q)=2\pi e^2/q$ is the two-dimensional Coulomb interaction in 
Fourier space and $\Pi({\bf q})$ is the static bare particle-hole propagator 
\begin{equation}
\Pi({\bf q})=-4\sum_{\lambda,\lambda',\textbf{k}}
|g_\textbf{k}^{\lambda,\lambda'}(\textbf{q})|^2
\frac{f[E_{\textbf{k}+\textbf{q}}^{\lambda'}]-
f[E_{\textbf{k}}^{\lambda}]}{E_{\textbf{k}+\textbf{q}}^{\lambda'}-
E_{\textbf{k}}^{\lambda}+i\eta}.
\end{equation}
Here, the factor of 4 takes into account of the spin and valley degeneracy, 
the index $\lambda =\pm 1$ denotes the two bands, and the vertex factor is 
given by \cite{wang2006},
\begin{equation}
\begin{split}
|g_\textbf{k}^{\lambda, \lambda}(\textbf{q})|^2 = \frac{1}{2}( 1
+\frac{\lambda \lambda' \sin \alpha_{\textbf{k}+\textbf{q}} \sin
 \alpha_\textbf{k} (k+q \cos
\theta)}{|\textbf{k}+\textbf{q}|} \\ +\lambda \lambda' \cos
\alpha_{\textbf{k} + \textbf{q}} \cos \alpha_\textbf{k} ),
\end{split}
\end{equation}
where $\theta$ is the angle between $\textbf{k}$ and $\textbf{q}$.
 $\alpha_\textbf{k}$ is defined through, $\tan \alpha_\textbf{k} = \frac{\hbar v_G
k}{\Delta_{so}}$. $E^\pm_\textbf{k}= \pm
\sqrt{\Delta^2_{so}+\hbar^2v_G^2k^2}$ is the energy spectrum of
the carriers in $\pm$ bands and $f[E_\textbf{k}^\lambda]$ is the Fermi
distribution function.

\begin {figure}[htpb]
\vskip 0.15cm
\includegraphics*[width= \linewidth]{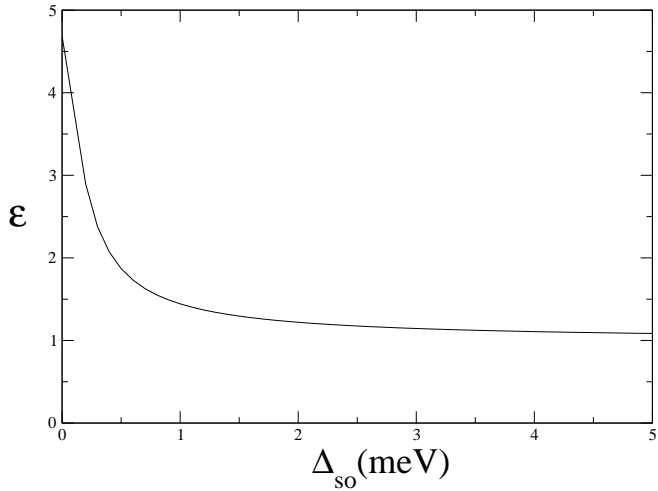}
\caption{Static dielectric constant $\epsilon(q\rightarrow 0,\omega=0)$ 
as a function of the spin-orbit interaction energy $\Delta_{so}$. We 
have used $\hbar v_G = 5.8$ eV\AA. We find that at $\Delta_{so} =4$ meV, 
$\epsilon \simeq 1.1$ approaches the minimum value possible $\epsilon=1$.}
\end{figure}

At low temperatures $T\rightarrow 0$, only the interband scattering contributes
to $\epsilon(\textbf{q})$. Figure 1 shows the dependence of $\epsilon$ on 
$\Delta_{so}$ evaluated numerically. We see that the dielectric constant 
decreases sharply with an increase in $\Delta_{so}$. When the spin orbit 
interaction energy is zero, the dielectric constant is $\epsilon=4.7$, 
whereas, $\epsilon \simeq 1.1$ for $\Delta_{so}=4$ meV. The decrease in 
dielectric constant increases the energy ratio to $E_p/E_k\simeq 1.4$. Even 
in the extreme case, $\epsilon=1$ (no screening whatsoever), the maximum value 
of the ratio is $E_p/E_k=1.5$, which is smaller than the ratio 
$E_p/E_k=r_s\sim 37$ required for Wigner crystallization in a conventional 2DEG. 

The second parameter $v_G$ can be tuned by putting strain on the system. 
In a tight-binding model, the velocity $v_G$ is related to the tunneling 
amplitude $t$ and the lattice constant $a$ by $v_G=3ta/2$. By straining the 
sample, the lattice constant increases linearly whereas the 
tunneling amplitude decreases exponentially, thus effectively decreasing 
the velocity of massless carriers. How effective this method will be in 
increasing the ratio $E_p/E_k$, depends upon the elastic properties of 
graphene; however, it seems unlikely that the velocity $v_G$ can change 
by an order of magnitude that will be required for Wigner crystallization. 

{\it Therefore, we conclude that Wigner crystal phase is absent 
in graphene}. The kinetic energy is always comparable to the potential 
energy, and both scale as $n^{3/2}$ with the carrier density. This 
peculiar dependence arises because of the linear dispersion of carriers. 

We note that these arguments will not be valid in the presence of a strong 
magnetic field. In that case, the system develops Landau levels with 
energies $E_m=\pm \hbar v_G/l_B\sqrt{2m}$ where $l_B$ is the magnetic length. 
Each Landau level is macroscopically degenerate. Therefore at small filling 
factors $\nu\ll 1$, we expect the 
carriers to undergo Wigner crystallization as there is no concommitant 
increase in the kinetic energy cost associated with the localization (This 
conclusion is robust for the $n=0$ Landau level, where the eigenfunctions 
are identical to those in the lowest Landau level of a conventional 2DEG). 
The critical value of the filling factor $\nu_c$ at which the Wigner 
crystallization occurs will have to be determined by a microscopic 
calculation, and is the subject of future work.  

This work was supported by DOE BES and LDRD at Los Alamos. We are grateful to
T. Wehling, M. Katsnelson, A. Lichtenstein, and R. Barnett for valuable discussion and
advice.

\bibliography{references}
\end{document}